\documentclass[aps,prl,twocolumn,showpacs,groupedaddress]{revtex4}  
\usepackage{graphicx}  
\usepackage{dcolumn}   
\usepackage{bm}        
\usepackage{amssymb}   

\begin{document}


\title{Bell-like inequality for spin-orbit separability of a classical laser beam}
\author{C.V.S.~Borges$^{1}$}
\author{M.~Hor-Meyll$^{1}$}
\author{J.A.O.~Huguenin$^{2}$}
\author{A.Z.~Khoury$^{1}$}

\affiliation{$^{1}$Universidade Federal Fluminense, Niter\'oi, Brazil}
\affiliation{$^{2}$Universidade Federal Fluminense, Volta Redonda, Brazil}
\date{\today}

\begin{abstract}
In analogy with Bell's inequality for two-qubit quantum states we propose an inequality
criterion for the non-separability of the spin-orbit degrees of freedom of a
\textit{classical} laser beam. A definition of separable and non-separable spin-orbit
modes is used in consonance with the one presented in Phys. Rev. Lett. \textbf{99},
160401 (2007). As the usual Bell's inequality can be violated for entangled two-qubit
quantum states, we show both theoretically and experimentally that the proposed
spin-orbit inequality criterion can be violated for non- separable modes. A discussion on
the classical-quantum transition is also presented.
\end{abstract}

 \maketitle

Experiments to show violation of Bell-like inequalities have attracted much attention in
the last years due to the possibility of ruling out classical hidden-variables theories
which jeopardize the need of a quantum mechanical model to describe nature. The majority
of proposed experiments relies on a \textit{pair} of entangled quantum particles for
genuine non-locality tests \cite{nonlocality1} or not necessarily entangled in the case
of non-contextuality tests \cite{contextuality1}.

Entanglement in \textit{single} particle degrees of freedom has already been investigated
in ref.\cite{contextuality2}, where a Bell-like inequality was violated by entangling the
spin and the beam path of single neutrons in an interferometer. The same kind of single
particle scheme has been proposed for photonic setups using the polarization and
transverse (spin-orbit) modes \cite{Spinorbit}. We have proposed a similar setup to
investigate the spin-orbit separability of a \textit{classical} laser beam in
\cite{QOIVParaty}. Simulations of Bell-inequalities in classical optics have also been
discussed in waveguides \cite{numericalsimulation} and imaging systems
\cite{OpticalClassical}. In this work we present our experimental results together with
the theoretical background developed for the analogy between the usual quantum mechanical
context of Bell inequality and our classical spin-orbit counterpart. Moreover, we discuss
the classical-quantum transition by assuming an initial coherent state of the
non-separable spin-orbit mode. We show that no genuine entanglement is present in the
classical optics implementation as expected. However, entanglement can be reached by post
selection of the single photon components of the initial coherent state.

Following ref. \cite{Topological}, we define as separable those spin-orbit modes that can
be written in the form $\vec{E}_{S}(\vec{r})=\psi(\vec{r})\hat{\mathbf{e}}$, where
$\psi(\vec{r})$ is a normalized c-number function of the transverse spatial coordinates
(transverse mode) and $\hat{\mathbf{e}}$ is a normalized polarization vector. However,
there are modes that cannot be written in this form, which we shall refer to as
non-separable. For example, consider the following normalized mode
\begin{eqnarray}
\vec{E}_{MNS}(\vec{r})=\frac{1}{\sqrt{2}}\left[\psi_{V}(\vec{r})\hat{\mathbf{e}}_{V}+
\psi_{H}(\vec{r})\hat{\mathbf{e}}_{H}\right]\;, \label{Efieldpsi}
\end{eqnarray}
where $\psi_{H}(\vec{r})$ and $\psi_{V}(\vec{r})$ are the first order Hermite-Gaussian
transverse modes with horizontal (H) and vertical (V) orientations \cite{Yariv}, and
$\hat{\mathbf{e}}_{H}$ and $\hat{\mathbf{e}}_{V}$ are the horizontal (H) and vertical (V)
linear polarization unit vectors. This mode cannot be written as product of a spatial
part times a polarization vector. In the space of spin-orbit modes of a classical beam,
it plays a role analogous to a maximally entangled two-qubit state, and we shall refer to
it as a maximally non-separable mode (MNS). While the separable spin-orbit modes have a
single polarization state over the beam wavefront, the non-separable modes exhibit a
polarization gradient leading to a polarization-vortex behavior \cite{polvortices}.

We can define an arbitrary classical spin-orbit mode as:
\begin{eqnarray}
\vec{E}(\vec{r})&=&A_{1}\psi_{V}(\vec{r})\hat{\mathbf{e}}_{V}+A_{2}\psi_{V}(\vec{r})\hat{\mathbf{e}}_{H}
+A_{3}\psi_{H}(\vec{r})\hat{\mathbf{e}}_{V}+\nonumber\\
& &A_{4}\psi_{H}(\vec{r})\hat{\mathbf{e}}_{H}, \label{Efieldarbitrary}
\end{eqnarray}
and discuss its separability with the aid of a concurrence-like quantity
\cite{Wooters,Topological}:
\begin{eqnarray}
C=2|A_{2}A_{3}-A_{1}A_{4}|
\end{eqnarray}
where $A_{i}$ ($i=1$ ... $4$) are complex numbers satisfying
$\sum_{i=1}^{4}|A_{i}|^{2}=1$. It turns out that $0<C\leq1$ for non-separable modes. In
particular we say that $C=1$ corresponds to a maximally non-separable mode. It can be
easily verified that any separable mode of the form:
\begin{eqnarray}
\vec{E}_{S}(\vec{r})=[B_{1}\psi_{V}(\vec{r})+B_{2}\psi_{H}(\vec{r})]
(B_{3}\hat{\mathbf{e}}_{V}+B_{4}\hat{\mathbf{e}}_{H})\;, \label{Efieldseparable}
\end{eqnarray}
where $B_{i}$ are arbitrary complex coefficients, has $C=0$.

To develop the spin-orbit inequality it will be useful to define the following rotated
basis of polarization and transverse modes:
\begin{eqnarray}
\hat{\mathbf{e}}_{\alpha+}&=&\cos{(2\alpha)}\hat{\mathbf{e}}_{V}+\sin{(2\alpha)}\hat{\mathbf{e}}_{H},\nonumber\\
\hat{\mathbf{e}}_{\alpha-}&=&\sin{(2\alpha)}\hat{\mathbf{e}}_{V}-\cos{(2\alpha)}\hat{\mathbf{e}}_{H}, \nonumber\\
\psi_{\beta+}(\vec{r})&=&\cos{(2\beta)}\psi_{V}(\vec{r})+\sin{(2\beta)}\psi_{H}(\vec{r}),\nonumber\\
\psi_{\beta-}(\vec{r})&=&\sin{(2\beta)}\psi_{V}(\vec{r})-\cos{(2\beta)}\psi_{H}(\vec{r}).
\label{Ebasis}
\end{eqnarray}

Re-writing the maximally non-separable mode given by Eq. (\ref{Efieldpsi}) in the rotated
basis we get:
\begin{eqnarray}
\vec{E}_{MNS}(\vec{r})&=&A_{e}(\psi_{\beta+}(\vec{r})\hat{\mathbf{e}}_{\alpha+}+\psi_{\beta-}(\vec{r})\hat{\mathbf{e}}_{\alpha-})+\nonumber\\
&
&A_{o}(\psi_{\beta-}(\vec{r})\hat{\mathbf{e}}_{\alpha+}-\psi_{\beta+}(\vec{r})\hat{\mathbf{e}}_{\alpha-})\;,
\label{Efieldpsinewbasis}
\end{eqnarray}
 where
\begin{eqnarray}
A_{e}&=&\cos{(2\alpha)}\cos{(2\beta)}+\sin{(2\alpha)}\sin{(2\beta)}\;, \nonumber\\
A_{o}&=&\cos{(2\alpha)}\sin{(2\beta)}-\sin{(2\alpha)}\cos{(2\beta)}\;.
\end{eqnarray}

Let $I_{\pm\pm}(\alpha,\beta)$ be the squared amplitude of the
$\psi_{\beta\pm}(\vec{r})\hat{\mathbf{e}}_{\alpha\pm}$ component in the expansion of
$\vec{E}_{MNS}(\vec{r})$ in the rotated basis. They play the same role as the detection
probabilities in the quantum mechanical context. Due to the orthonormality of
$\{\psi_{\beta+}, \psi_{\beta-}\}$ and $\{\hat{\mathbf{e}}_{\alpha+},
\hat{\mathbf{e}}_{\alpha-}\}$ it can be easily shown that:
\begin{eqnarray}
I_{++}(\alpha,\beta)+I_{+-}(\alpha,\beta)+I_{-+}(\alpha,\beta)+I_{--}(\alpha,\beta)=1\;.
\end{eqnarray}
Following the analogy with the usual quantum mechanical Bell inequality for spin $1/2$
particles, we can define
\begin{eqnarray}
M(\alpha,\beta)&=&I_{++}(\alpha,\beta)+I_{--}(\alpha,\beta)-I_{+-}(\alpha,\beta)
\nonumber\\
&-&I_{-+}(\alpha,\beta)\;,
\end{eqnarray}
and derive a Bell-type inequality for the quantity
\begin{eqnarray}
S = M(\alpha_{1},\beta_{1})+M(\alpha_{1},\beta_{2})-M(\alpha_{2},\beta_{1})+M(\alpha_{2},\beta_{2})\;.\nonumber\\
\label{inequality}
\end{eqnarray}
For any separable mode, $-2\leqslant S \leqslant2$, however this condition can be
violated for non-separable modes. A maximal violation of the previous inequality,
corresponding to $S=2\sqrt{2}$, can be obtained for the set:
\begin{eqnarray}
\alpha_{1}=\pi/16\,,\; \alpha_{2}=3\pi/16\,,\; \beta_{1}=0\,,\; \beta_{2}=\pi/8\;.
\label{basis}
\end{eqnarray}

\begin{figure}
\includegraphics[scale=0.45]{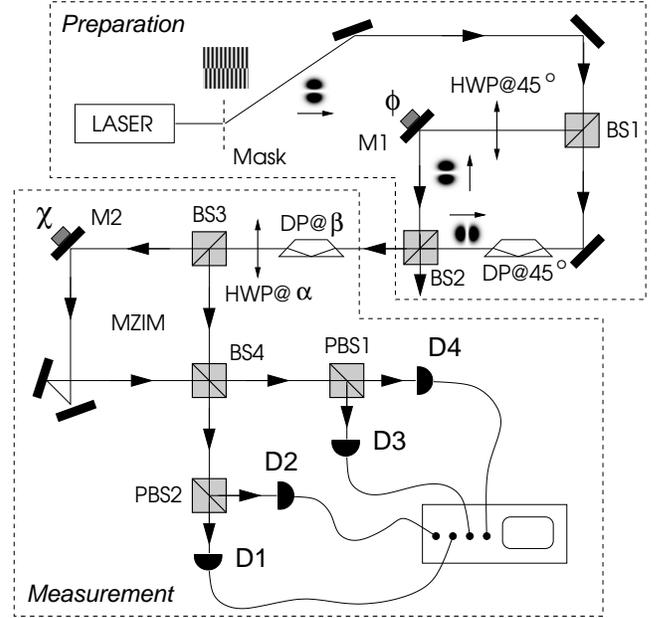}
\caption{\label{fig:setup} Experimental setup for the Bell-type inequality violation
using a non-separable classical beam. HWP - half-wave plate, DP - Dove prism, (P)BS -
(polarizing) beam splitter, D1, D2, D3, and D4 - photo-current detectors.}
\end{figure}

The experimental setup to observe the maximum violation of the non-separability
inequality is shown in Figure \ref{fig:setup}
and is composed by two stages: preparation of the maximally non-separable mode and
measurement of the intensities $I_{\pm\pm}(\alpha,\beta)$. The preparation stage consists
of a Mach-Zender (MZ) interferometer with a half-wave plate (HWP) oriented at
$45^{\circ}$ with respect to the horizontal plane in one arm and a Dove prism (DP) also
oriented at $45^{\circ}$ with respect to the horizontal plane in the other arm. Before
the MZ interferometer, a holographic mask positioned in the path of a horizontally
polarized TEM$_{00}$ laser beam produces mode
$\vec{E}(\vec{r})=\psi_{V}(\vec{r})\hat{\mathbf{e}}_{H}$ at the first diffraction order.
In the MZ interferometer, the half-wave plate converts $\hat{\mathbf{e}}_{H}$ into
$\hat{\mathbf{e}}_{V}$ and the Dove prism changes $\psi_{V}(\vec{r})$ into
$\psi_{H}(\vec{r})$ so the resulting mode at the output BS2 is:
\begin{eqnarray}
\vec{E}(\vec{r})= \frac{1}{\sqrt{2}}\left[\psi_{H}(\vec{r})\hat{\mathbf{e}}_{H}+
e^{i\phi}\psi_{V}(\vec{r})\hat{\mathbf{e}}_{V}\right]\;, \label{Eprepared}
\end{eqnarray}
where $\phi$ is the phase difference between the two arms of the MZ interferometer.
Mirror M1 is mounted on a piezoelectric transducer (PZT) to allow fine control of the
phase difference $\phi$. Our goal is to prepare the initial mode with $\phi=0$, but we
carry out this phase in the calculations to show how our results depends on it. The other
output of BS2 is used to check the alignment between the two components of the mode
prepared.

The measurement stage is composed by a Dove prism oriented at a variable angle $\beta$
(DP@$\beta$), a half-wave plate oriented at a variable angle $\alpha$ (HWP@$\alpha$), a
Mach-Zender interferometer with an additional mirror (MZIM) \cite{MZIM}, and one
polarizing beam splitter (PBS) after each of the MZIM outputs. Four photo-current
detectors (D1, D2, D3 and D4) are used to measure the intensities at the PBS outputs.
HWP@$\alpha$ combined with DP@$\beta$ define in which basis we are going to measure our
initial mode.

We want MZIM to work as a parity selector delivering odd modes
$\psi_{V}(\vec{r})\hat{\mathbf{e}}_{H}$ and $\psi_{H}(\vec{r})\hat{\mathbf{e}}_{V}$ in
one port and even modes $\psi_{V}(\vec{r})\hat{\mathbf{e}}_{V}$ and
$\psi_{H}(\vec{r})\hat{\mathbf{e}}_{H}$ in the other port. Parity is evaluated according
to the eigenvalue of the respective mode under reflection over the horizontal plane. Let
$\chi$ be the optical phase difference between the two arms of the MZIM. Note that proper
functioning of the MZIM as a parity selector occurs only when $\chi = 2n\pi$
($n=0,1,2,$...). For $\chi = (2n+1)\pi$, the even and odd outputs are interchanged.

After propagating through DP@$\beta$ and through HWP@$\alpha$, the maximally
non-separable mode given by Eq. (\ref{Eprepared}) transforms to:
\begin{eqnarray}
\vec{E'}(\vec{r})&=&A^{++}(\phi)\psi_{V}(\vec{r})\hat{\mathbf{e}}_{V}+A^{+-}(\phi)\psi_{V}(\vec{r})\hat{\mathbf{e}}_{H}
\nonumber\\
&+&A^{-+}(\phi)\psi_{H}(\vec{r})\hat{\mathbf{e}}_{V}+A^{--}(\phi)\psi_{H}(\vec{r})\hat{\mathbf{e}}_{H},
\label{Epreparedbasis}
\end{eqnarray}
where
\begin{eqnarray}
 A^{++}(\phi)=e^{i\phi}\cos{(2\alpha)}\cos{(2\beta)}+\sin{(2\alpha)}\sin{(2\beta)}\;,\nonumber\\
 A^{+-}(\phi)=e^{i\phi}\sin{(2\alpha)}\cos{(2\beta)}-\cos{(2\alpha)}\sin{(2\beta)}\;,\nonumber\\
 A^{-+}(\phi)=e^{i\phi}\cos{(2\alpha)}\sin{(2\beta)}-\sin{(2\alpha)}\cos{(2\beta)}\;,\nonumber\\
 A^{--}(\phi)=e^{i\phi}\sin{(2\alpha)}\sin{(2\beta)}+\cos{(2\alpha)}\cos{(2\beta)}\;.\label{A++--}
\end{eqnarray}

If MZIM phase $\chi=0$ then the four amplitudes above would be the ones measured by the
detectors since MZIM interferometer together with PBS1 and PBS2 would separate the modes
$\psi_{V}(\vec{r})\hat{\mathbf{e}}_{V}$, $\psi_{V}(\vec{r})\hat{\mathbf{e}}_{H}$,
$\psi_{H}(\vec{r})\hat{\mathbf{e}}_{V}$ and $\psi_{H}(\vec{r})\hat{\mathbf{e}}_{H}$. But
we will still consider the case in which $\chi$ may differ from zero, and then the
corresponding intensities normalized to the total intensity are given by:

\begin{eqnarray}
I_{1}&=&\sin^{2}{(\chi/2)}|A^{--}(\phi)|^{2}+\cos^{2}{(\chi/2)}|A^{+-}(\phi)|^{2},\nonumber\\
I_{2}&=&\sin^{2}{(\chi/2)}|A^{++}(\phi)|^{2}+\cos^{2}{(\chi/2)}|A^{-+}(\phi)|^{2},\nonumber\\
I_{3}&=&\cos^{2}{(\chi/2)}|A^{++}(\phi)|^{2}+\sin^{2}{(\chi/2)}|A^{-+}(\phi)|^{2},\nonumber\\
I_{4}&=&\cos^{2}{(\chi/2)}|A^{--}(\phi)|^{2}+\sin^{2}{(\chi/2)}|A^{+-}(\phi)|^{2}.
\label{IHHVV}
\end{eqnarray}

We can test the violation of the non-separability inequality by making measurements in
the bases $(\alpha_1,\beta_1)$, $(\alpha_1,\beta_2)$, $(\alpha_2,\beta_1)$ and
$(\alpha_2,\beta_2)$ and obtaining the values of $M(\alpha,\beta)$ and subsequently $S$.
The value of $S$ for arbitrary $\phi$ and $\chi$ is given by:
\begin{eqnarray}
S(\chi,\phi)=\sqrt{2}\cos{\chi}(1+\cos{\phi}). \label{Sexp}
\end{eqnarray}
Thus maximal violation of the non-separability inequality is accomplished for
$\phi=\chi=0$. This is a key result because it shows that experimental errors in the
phases will only diminish the violation, not increase it.

\begin{figure}
\includegraphics[scale=0.3]{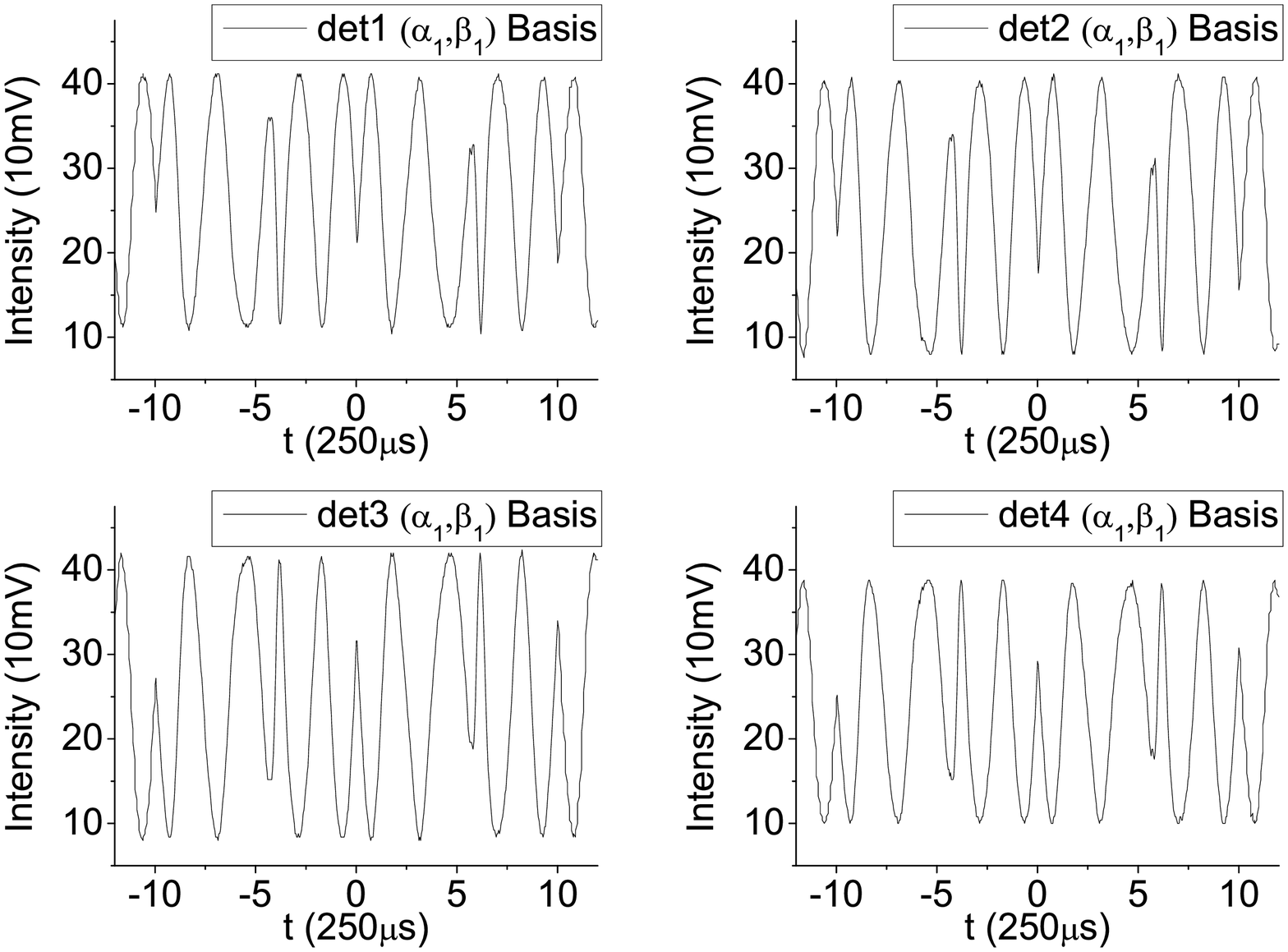}
\caption{\label{fig:graphics} Experimental results for the maximally non-separable
initial mode, measured in the $(\alpha_1,\beta_1)$ basis. Time parametrizes the MZIM
phase $\chi$.}
\end{figure}

In our experiment the MZIM phase $\chi$ is continuously varied by applying a voltage ramp
to the PZT on M2 while intensities $I_1$ through $I_4$ are monitored at the oscilloscope.
An example of our experimental results is presented in Fig. \ref{fig:graphics}, showing
the oscillations caused by the variation of $\chi$. We know from the intensities
dependence on $\phi$ and $\chi$ that $\chi=0$ corresponds to the peaks in the graphics,
and $\phi=0$ corresponds to a maximal visibility of these oscillations. Since we have a
few repetitions of these peaks, we obtain an ensemble of intensities which allows us to
calculate the average of $M(\alpha,\beta)$ and $S$, whose values are show in Table
\ref{tab:averages}. The standard deviations for the MNS mode results is within $2\%$. In
this table we also show our experimental results for a separable initial mode
$\psi_V(\vec{r})\hat{\mathbf{e}}_V$, which is easily obtained by blocking the Dove Prism
arm of the preparation MZ interferometer.

In Table \ref{tab:table} we show our best experimental results for the intensities $I_1$
through $I_4$ which lead to a violation with $S=2,17$.

\begin{table}
\caption{\label{tab:averages}Mean values for $M$ and $S$ for maximally non-separable and
separable modes.}
\begin{ruledtabular}
\begin{tabular}{cccccc}
mode & $\bar{M}(\alpha_1,\beta_1)$ & $\bar{M}(\alpha_1,\beta_2)$ & $\bar{M}(\alpha_2,\beta_1)$ & $\bar{M}(\alpha_2,\beta_2)$ & $\bar{S}$ \\
MNS & $0,615$ & $0,49$ & $-0,525$ & $0,49$ & $2,11$ \\
S & $0,47$ & $0,00$ & $-0,56$ & $0,00$ & $1,03$ \\
\end{tabular}
\end{ruledtabular}
\end{table}

\begin{table}
\caption{\label{tab:table}Best experimental data for maximally non-separable initial mode
(Intensities are given in 10mV).}
\begin{ruledtabular}
\begin{tabular}{ccccccc}
basis&$I_{1}$&$I_{2}$&$I_{3}$&$I_{4}$&$I_{tot}$&$M$\\
\hline $(\alpha_1, \beta_1)$ & $7,99$ & $10,4$ & $41,6$ & $38,4$ & $98,39$ & $0,626$\\
$(\alpha_1, \beta_2)$ & $13,6$ & $10,4$ & $30,8$ & $43,6$ & $98,4$ & $0,512$\\
$(\alpha_2, \beta_1)$ & $36,4$ & $40,4$ & $10,8$ & $12,8$ & $100,4$ & $-0,530$\\
$(\alpha_2, \beta_2)$ & $13,2$ & $11,6$ & $29,6$ & $45,2$ & $99,6$ & $0,502$\\
\end{tabular}
\end{ruledtabular}
\end{table}

Attenuation of the intense laser beam down to the photon count regime brings the setup to
the quantum mechanical domain. In order to discuss the classical-quantum transition of
this experiment, let us first use a coherent state to represent the intense laser beam
prepared in the maximally non-separable mode
\begin{eqnarray}
|\alpha\rangle_{MNS}=
e^{\frac{-|\alpha|^{2}}{2}}\sum_{n=0}^{\infty}\frac{\alpha^{n}(a^{\dagger}_{MNS})^n}{n!}\;|0\rangle\;,
\label{coherentstate}
\end{eqnarray}
where $a^{\dagger}_{MNS}$ is the creation operator associated with the MNS mode. Its
action on the vacuum state $|0\rangle$ produces a one-photon Fock state in the MNS mode.
Since this mode is decomposed as in eq.(\ref{Efieldpsi}), its corresponding creation
operator can be written as
$a^{\dagger}_{MNS}=(a^{\dagger}_{VV}+a^{\dagger}_{HH})/\sqrt{2}$, where the first index
corresponds to the transverse mode and the second to the polarization mode, so that

\begin{eqnarray}
|\alpha\rangle_{MNS}&=&
e^{\frac{-|\alpha|^{2}}{2}}\sum_{n=0}^{\infty}\sum_{q=0}^{n}\left(\frac{\alpha}{\sqrt{2}}\right)^{q}
\frac{(a^{\dagger}_{VV})^q}{q!}\nonumber\\
&\times&\left(\frac{\alpha}{\sqrt{2}}\right)^{n-q}
\frac{(a^{\dagger}_{HH})^{n-q}}{(n-q)!}\;|0\rangle\;. \label{coherentstate2}
\end{eqnarray}

Exchanging the order of the summations and defining $m\equiv n-q$, we obtain
\begin{eqnarray}
|\alpha\rangle_{MNS}&=&e^{\frac{-(|\alpha|/\sqrt{2})^{2}}{2}}\sum_{q=0}^{\infty}\left(\frac{\alpha}{\sqrt{2}}\right)^{q}
\frac{(a^{\dagger}_{VV})^q}{q!}\nonumber\\
&\times&e^{\frac{-(|\alpha|/\sqrt{2})^{2}}{2}}\sum_{m=0}^{\infty}\left(\frac{\alpha}{\sqrt{2}}\right)^{m}
\frac{(a^{\dagger}_{HH})^m}{m!}\;|0\rangle\nonumber\\
&=&|\alpha/\sqrt{2}\rangle_{VV}|\alpha/\sqrt{2}\rangle_{HH}\;,
\end{eqnarray}
which clearly shows that it is a product of coherent states at modes $HH$ and $VV$ with
complex amplitude $\alpha/\sqrt{2}$.

However, it is instructive to take a look at the Fock state decomposition of
$|\alpha\rangle_{MNS}$:
\begin{eqnarray}
|\alpha\rangle_{MNS}=e^{\frac{-|\alpha|^{2}}{2}}[|0\rangle+
\frac{\alpha}{\sqrt{2}}(|1_{HH}0_{VV}\rangle+|0_{HH}1_{VV}\rangle)+...]\nonumber\\
\end{eqnarray}
The single photon component is clearly a maximally entangled state. Therefore,
post-selection of single photon states from $|\alpha\rangle_{MNS}$ followed by the
experimental setup used here allows one to investigate the usual Bell inequality from
probability measurements.

In conclusion, we have investigated both theoretically and experimentally an inequality
criterion, as a sufficient condition, for the spin-orbit non-separability of a classical
laser beam. The notion of separable and non-separable spin-orbit modes in classical
optics builds an useful analogy with entangled quantum states, allowing for the study of
some of their important mathematical properties. This analogy has already been
successfully exploited in our group to investigate the topological nature of the phase
evolution of an entangled state under local unitary operations \cite{Topological}. Many
quantum computing tasks require entanglement but do not need nonlocality, so that using
different degrees of freedom of single particles can be useful. This is the type of
entanglement whose properties can be studied in the classical optical regime allowing one
to replace time-consuming measurements based on photon count by the much more efficient
measurement of photocurrents.

Although helpful, the notion of mode non-separability must not be confused with genuine
quantum entanglement. In order to avoid this confusion, we have included a brief
discussion of the classical-quantum transition. We show that a coherent quantum state of
a maximally non-separable mode can be written as the product of two coherent states in
the separable components of the MNS mode. On the other hand, its Fock state decomposition
shows that the single photon component exhibits entanglement that can be accessed through
post-selection.

We would like to thank  Luiz Davidovich and Adriana Auyuanet for discussions during the
development of this work. The authors acknowledge financial support from the Brazilian
funding agencies CNPq, CAPES and FAPERJ.  This work was performed as part of the
Instituto Nacional de Ci\^encia e Tecnologia de Informa\c c\~ao Qu\^antica (CNPq
funding).

\end{document}